\documentclass[11pt]{article}
\newcommand{\tab}{\hspace*{2em}}

\usepackage[preprint]{acl}

\usepackage{times}
\usepackage{latexsym}

\usepackage[T1]{fontenc}

\usepackage[utf8]{inputenc}

\usepackage{microtype}

\usepackage{inconsolata}

\usepackage{graphicx}
\usepackage{booktabs}

\title{Tool-Guided Retrieval-Augmented Repair for Securing LLM-Generated C Code}

\author{Vidyut Sriram \\
  The Pennsylvania State University \\
  \texttt{vps5353@psu.edu} \\\And
  Saatvik Pradhan \\
  The Pennsylvania State University \\
  \texttt{szp6118@psu.edu} \\
  \AND
  Suman Saha \\
  The Pennsylvania State University \\
  \texttt{szs339@psu.edu} \\}

\begin{document}
\maketitle
\begin{abstract}
\tab Large language models can generate C code from natural-language descriptions, but resulting programs often contain security vulnerabilities and compilation errors, posing risks for embedded and resource-constrained systems. This work investigates how feedback and retrieval improve reliability of LLM-generated C code. We present an analysis-and-repair workflow that combines compilation diagnostics, CodeQL static analysis, and KLEE symbolic execution with retrieval of prior repair patterns for iterative refinement.

Evaluated on 5,000 C programming tasks exercising embedded relevant vulnerabilities, baseline models show substantial reliability gaps, with compilation failure rates up to 46\% and security defect rates up to 49\%. Our approach improves both metrics. For CodeLlama 7B, security defect rates decrease from 49\% to 19\% and total CodeQL errors drop from 15,088 to 2,463 (83.7\%). For DeepSeek Coder 1.3B, compilation failures are reduced from 42\% to 22\% and security defects from 35\% to 15\%. These results show that integrating lightweight analysis tools can improve the safety of LLM-generated code for embedded development.

\end{abstract}

\section{Introduction}
Large language models (LLMs) are increasingly used to assist low-level programming tasks, including firmware and systems code in C. While they generate syntactically plausible programs from natural-language descriptions, outputs often contain security vulnerabilities and unsafe patterns. Prior work shows that GitHub Copilot can produce vulnerable code in security-relevant cases \cite{pearce2022}, and that developers using AI code assistants may write less secure code than those without such tools \cite{perry2023}. More recent studies confirm that modern code-generation models still exhibit nontrivial vulnerability rates on secure coding benchmarks \cite{yan2025}. These limitations raise concerns about deploying LLM-generated code in safety-critical and resource-constrained environments.

The risks are particularly acute in embedded systems. Firmware often runs on resource-constrained platforms with limited memory, architecture-specific constraints, and fewer practical protections than general-purpose systems. In such settings, memory-safety violations, unchecked inputs, or misuse of low-level APIs can lead to persistent device compromise, denial-of-service, or costly field updates \cite{ulhaq2023survey}. As a result, relying on single-pass LLM code generation followed by manual inspection is not sufficient for this setting.

Recent research has explored two main directions for improving the security of LLM-generated code. Retrieval-augmented approaches guide generation using external security knowledge or examples, but naive retrieval from security documents can introduce noise and weaken guidance \cite{shi2025rescue}. A second line of work uses iterative repair driven by tool feedback, especially static analysis, to refine insecure code, showing gains on higher-level language benchmarks such as Python \cite{khoury2025fdsp}. However, these approaches remain focused on general-purpose settings and do not address the combined requirements of embedded C, where security, compilation correctness, and low-level behavior must be considered together. This gap motivates a workflow that coordinates multiple verification signals rather than relying on retrieval alone or a single feedback source.

This paper takes a step toward addressing this gap by proposing a tool-guided generation and repair pipeline for C code that combines compilation feedback, static security analysis, and a final semantic validation stage. The core idea is to prioritize repairs by impact, addressing security issues before compilation failures, while using an accumulated repository of prior repair outcomes to guide later generations. This design reflects how developers reason about low-level code, where eliminating unsafe behavior is often more important than producing compilable but insecure programs. Unlike prior work that treats retrieval and repair as separate strategies, this approach connects them through a shared memory of repair patterns, enabling reuse of validated guidance across tasks.

We evaluate this approach on a dataset of 5,000 C programming tasks using three open-source models. Preliminary results show substantial reductions in compilation failures and static-analysis-detected security issues, particularly for the smaller model. These findings suggest that structured, tool-guided feedback can improve the reliability of LLM-generated C code without requiring larger models or retraining. As a work in progress, this paper demonstrates the feasibility of such a workflow, showing that coordinated compilation and static analysis feedback reduce both compilation failures and security defects on general C tasks. Future work will extend evaluation to embedded benchmarks, analyze improvements across vulnerability classes, and conduct ablation studies to better understand component contributions and real-world applicability.
\section{Related Work}

Early work on secure code generation mainly established the problem rather than solving it. Pearce et al.~\cite{pearce2022} showed that GitHub Copilot can generate vulnerable code, while Perry et al.~\cite{perry2023} found that developers using AI coding assistants may produce less secure code than those working without such assistance. These studies motivate the need for methods that improve the security of LLM-generated code, but they do not propose a repair workflow for low-level C code.

Subsequent work has explored several strategies for mitigating these weaknesses. SecCoder improves secure code generation through carefully selected demonstrations and dense retrieval, showing that external secure examples can help guide models toward safer outputs~\cite{seccoder2024}. A different line of work uses iterative refinement driven by security tools. FDSP incorporates static-analysis feedback into a closed-loop repair process and reports improved security on Python benchmarks~\cite{khoury2025fdsp}. SecureFixAgent follows a similar detect-repair-validate pattern using a lightweight local model and Bandit for Python vulnerability repair~\cite{securefixagent2025}. Together, these studies show that both retrieval and tool-guided refinement can improve secure code generation.

Among prior methods, retrieval-based secure generation is the closest to our approach. SecCoder uses retrieval to provide secure demonstrations~\cite{seccoder2024}, and RESCUE further shows that naive retrieval from raw security documents can introduce noise, motivating more selective security-focused retrieval~\cite{shi2025rescue}. These methods demonstrate the value of retrieval, but they primarily use retrieval to improve generation quality rather than to support an incremental repair process grounded in prior repair outcomes.

Our work differs from prior approaches in setting and design. Existing methods either use retrieval to supply secure examples or apply repair in higher-level languages with a single feedback source. In contrast, we target C code in an embedded-oriented workflow and combine retrieval with iterative repair, integrating compiler and static analysis feedback with final-stage symbolic validation. Our approach also uses a repository of prior repair outcomes, enabling reuse across tasks and coordinating multiple verification signals.
\section{Approach}
Our approach proceeds in four stages: (1) task specification and initial generation, (2) compilation and CodeQL-based static security analysis, (3) retrieval-guided repair using a repository of verified repair patterns, and (4) one-shot symbolic validation with KLEE.

\subsection{Task Specification and Initial Generation}
Each input task specifies a C function or small program from a dataset of algorithmic problems, including mathematical computation, string processing, and sorting. The model prompt includes the natural-language task description and a short security checklist. We do not provide extra type signatures, platform details, or hardware constraints beyond standard C compilation. Given this input, the model generates an initial candidate implementation, passed directly to the analysis stage without human intervention.

\subsection{Compilation and Static Analysis}
The first analysis stage evaluates whether the generated code is buildable and contains statically detectable security issues. We begin by compiling the candidate using GCC. If compilation fails, the resulting diagnostics, such as missing headers, undeclared identifiers, or type mismatches, are collected and summarized for use in later repair prompts. If compilation succeeds, we run static security analysis on the same candidate using CodeQL, a semantic code analysis framework, with queries targeting vulnerability classes common in C programs. These include buffer overflows and out-of-bounds memory accesses, use of unsafe library functions such as \texttt{gets}, \texttt{strcpy}, and unbounded \texttt{sprintf}, unchecked return values from I/O and memory operations, and integer-related issues such as overflow and signedness errors. For each finding, we record the corresponding rule identifier and diagnostic message (including file location when available). Together, compilation diagnostics and CodeQL findings provide the structured feedback used during repair in the next stage.

\subsection{Retrieval-Augmented Repair Patterns}
A central component of the approach is an execution-time repository of repair patterns that grows across tasks. Rather than relying on a fixed external knowledge base, the repository updates incrementally: after each task attempt, the candidate and analysis outcomes are stored and used to guide later generations. As more tasks are processed, the retrieval context improves by accumulating verified repair behaviors.

Each repository entry includes (i) the natural-language task description, (ii) the generated C code, and (iii) metadata summarizing analysis outcomes such as compilation status, CodeQL security status, KLEE analyzability, raw CodeQL findings, and a composite quality score. The quality score determines retrieval priority and final selection. Non-compiling candidates receive a score of 0.0. For compiling candidates, the score is 0.25 if the code only compiles, 0.50 if it compiles and is KLEE-analyzable but still has CodeQL findings, 0.75 if it compiles and is CodeQL-clean, and 1.0 if it compiles, is CodeQL-clean, and KLEE-clean. The scheme prioritizes security cleanliness while distinguishing partial success.

For a new task, the repository is queried using the task description to retrieve semantically similar prior entries. Retrieval follows a contrastive design. Up to three high-quality entries with a quality score above 0.60 (compiling and CodeQL-clean) are used as positive evidence. Up to two insecure but compilable entries are used as negative evidence, and only when at least three positive examples are available, so that avoidance guidance does not dominate the prompt.

Retrieved entries are not inserted as few-shot demonstrations. Instead, the system distills them into brief textual guidance. Positive examples yield up to seven security practices from prior successful code, such as checking \texttt{scanf} return values, validating memory allocation, and using bounded string operations. Negative examples provide up to three avoidance hints derived from CodeQL findings. These practices and warnings are given as concise instructions rather than raw code, reducing the risk that smaller models copy examples without adapting them to the current task.

If a candidate shows compilation failures or security findings, the system triggers a bounded repair process. For each task, it generates an initial candidate and performs up to three repair iterations, producing at most four programs. Each iteration’s prompt includes the task, code, diagnostics, and retrieval-derived guidance. The model outputs a revised candidate, which is recompiled and reanalyzed. After all attempts, the best version is chosen using the composite quality score. To preserve repository quality, we deduplicate near-identical entries by removing candidates with cosine similarity above 0.90 and keeping the higher-quality version. The repository is capped at 500 items by discarding the lowest-scoring patterns when over the limit.

\subsection{Final Symbolic Validation}
After the repair process, the final candidate undergoes symbolic execution with KLEE. This stage lies outside the repair loop and produces no additional prompts; instead, it serves as a one-shot semantic check to detect errors missed by static analysis. Symbolic execution explores feasible program paths using symbolic inputs, which can expose concrete failing behaviors (via counterexample inputs) such as assertion failures, out-of-bounds accesses, null dereferences, or other memory-safety violations. We use KLEE to assess whether the final candidate is analyzable under our harness assumptions and whether any violations are reachable. Regardless of outcome, we record the KLEE result (KLEE-analyzable and any violations found) in the repository metadata and update the composite quality score, so future retrieval prefers patterns that are both security-clean and symbolically robust.

\section{Experiment Results}

We present the experimental setup and key results for three models, evaluated with and without the proposed workflow. The evaluation uses a dataset of general-purpose C tasks that, while not embedded-specific, capture common vulnerability classes such as unchecked inputs, unbounded memory writes, and unsafe library usage, which are representative of typical failure modes in embedded C development.

We evaluate on a dataset of 5{,}000 C programming tasks drawn from general-purpose algorithmic problems, including numerical computation, sorting, and string processing. Each task provides a natural-language description and requires a compilable C program. The tasks cover common sources of security defects in C, including array and integer manipulation, memory use, string handling, and input parsing with \texttt{scanf}. They thus reflect vulnerability classes such as unchecked inputs, unbounded writes, unsafe library usage, and improper memory handling. Although not embedded-specific, these patterns mirror common failure modes in embedded C. We therefore use this dataset as a proxy for reliability evaluation, leaving direct embedded tests for future work. We assess DeepSeek Coder 1.3B and CodeLlama 7B using identical prompts and greedy decoding (do\_sample=False), so outputs are deterministic per task–model pair. For each task, the model generates one initial candidate and then performs up to three repair iterations, for at most four attempts.

\subsection{Baseline Reliability}
We assess model reliability using three metrics: compilation failure rate, static defect rate, and KLEE analyzability. Programs that fail to compile are excluded from static security checks since CodeQL is not applied to non-compiling code; these cases are reported separately rather than labeled security-clean, and this convention holds for both baseline and repair conditions. As shown in Table~\ref{tab:rag-summary}, baseline reliability is limited across all models. DeepSeek Coder and CodeLlama have high compilation failure rates ($42.44\%$ and $45.56\%$), while Qwen2.5 performs better but still fails on over a quarter of tasks ($27.18\%$). Security issues are widespread, with CodeLlama showing the highest defect rate ($48.50\%$), followed by DeepSeek Coder ($34.62\%$) and Qwen2.5 ($27.90\%$). The most frequent CodeQL findings involve missing input validation and unsafe memory operations, such as unchecked \texttt{scanf} return values, unbounded format strings, and unsafe use of \texttt{strcpy} and \texttt{sprintf}. Additional issues include unchecked allocation sizes and potential heap overflows.

\begin{table*}[t]
  \centering
  \caption{Baseline vs.\ Triage-RAG results on C tasks.}
  \label{tab:rag-summary}
  
    \begin{tabular}{lccccc}
      \toprule
      Model & Setting & Comp.\ & Security & KLEE  & Total security  \\
      & & fail (\%) & issues (\%) & analyzable (\%) & errors \\
      \midrule
      DeepSeek 1.3B & Baseline   & 42.44 & 34.62 & 56.80 & 3{,}292 \\
                    & RAG-based & 21.78 & 15.16 & 77.44 & 1{,}870 \\
      \midrule
      CodeLlama 7B  & Baseline   & 45.56 & 48.50 & 54.10 & 15{,}088 \\
                    & RAG-based & 27.50 & 19.32 & 71.20 & 2{,}463 \\
      \bottomrule
    \end{tabular}%

\end{table*}

\subsection{RAG-Enhanced Repair Results}
Table~\ref{tab:rag-summary} also reports results with the repair workflow enabled. For both models, the workflow consistently reduces compilation failures and CodeQL-flagged issues while improving KLEE analyzability. The largest gain appears for CodeLlama 7B, where total CodeQL errors drop from 15{,}088 to 2{,}463 (83.7\% reduction). DeepSeek Coder 1.3B likewise improves, with security issues falling from 34.62\% to 15.16\% and KLEE analyzability rising from 56.80\% to 77.44\%. Because CodeQL does not run on non-compiling programs, we examine security-clean rates conditioned on successful compilation. Under this view, CodeLlama rises from 10.9\% to 73.4\%, while DeepSeek Coder climbs from 65.4\% to 80.6\%. These results indicate gains reflect higher build success and fewer security findings among compiled programs, not compilation alone.

\subsection{Top Error Types and NIST Taxonomy Alignment}
Static analysis of repaired outputs from CodeLlama and DeepSeek shows that remaining security issues align closely with established NIST software flaw categories.\footnote{\url{https://www.nist.gov/itl/ssd/software-quality-group/taxonomy-software-flaws}}
Overall, 60.2\% of CodeLlama's 2{,}463 CodeQL findings and 58.0\% of DeepSeek's 1{,}870 findings map to NIST-defined flaw classes. In both models, the most common NIST-aligned finding is missing checks on input operations such as \texttt{scanf}, under \emph{Exceptional Condition Handling}. This pattern reflects failure to validate return values and remains dominant after repair. DeepSeek also shows notable unbounded write findings mapping to the \emph{Buffer Overflow} category. Several remaining findings, including unused variables and implicit function declarations, do not map cleanly to NIST and are better treated as code-quality or portability issues rather than security flaws. These results suggest that even after retrieval-guided repair, LLM-generated C code retains well-known weaknesses, with \emph{Exceptional Condition Handling} and \emph{Buffer Overflow} remaining the most persistent NIST-aligned categories. Table~\ref{tab:nist-alignment} summarizes the NIST alignment.

\begin{table}[t]
  \centering
  \caption{NIST alignment of CodeQL findings after repair.}
  \label{tab:nist-alignment}

  \small
  \begin{tabular}{lccc}
    \toprule
    Model & NIST- & Non- & Top  \\
         & mapped & NIST & NIST Type \\
    \midrule
    CodeLlama 7B  & 60.2\% & 39.8\% & Ex.\ cond.\ (51.8\%) \\
    DeepSeek 1.3B & 58.0\% & 42.0\% & Ex.\ cond.\ (40.8\%) \\
                  &        &        & Buffer (12.2\%) \\
    \bottomrule
  \end{tabular}%
  
\end{table}
\section{Future Work}

In the full paper, we will extend this WIP in three concrete directions: First, we will evaluate the workflow on embedded-focused benchmarks, including firmware-style tasks, to test whether gains from general C programs transfer to low-level settings motivating this work. Second, we will replace aggregate error reporting with taxonomy-based vulnerability analysis and more interpretable symbolic-execution results, linking improvements to specific flaw types and reachable behavioral failures rather than raw warning counts. Third, we will add a lightweight ablation study to isolate the contribution of retrieval and multi-tool feedback, clarifying which components drive the observed improvements.
\section{Conclusion}

This work presents a tool-guided workflow for improving reliability and security of LLM-generated C programs. The approach combines retrieval from prior repair outcomes with feedback from compilation, static analysis, and symbolic execution to guide refinement. Evaluated on 5,000 C tasks, the method yields consistent gains across two models: CodeLlama 7B cuts security defect rates from 49\% to 19\% and total CodeQL errors from 15,088 to 2,463, while DeepSeek-Coder 1.3B lowers compilation failures from 42\% to 22\% and security defects from 35\% to 15\%. These findings show that incorporating lightweight analysis tools into the generation loop improves build success and security, with embedded evaluation and component ablations left for future work.

\bibliography{sample-base}
\end{document}